\documentclass[aps,showpacs]{revtex4}
\usepackage{amssymb}
\usepackage{graphicx}
\usepackage{dcolumn}
\usepackage{amsmath}

\begin{document}

\title{The Kondo-lattice state and non-Fermi-liquid behavior in the presence
of Van Hove singularities}
\author{V. Yu. Irkhin}
\email{Valentin.Irkhin@imp.uran.ru}
\affiliation{Institute of Metal Physics, 620041 Ekaterinburg, Russia}

\begin{abstract}
A scaling consideration of the Kondo lattices is performed with account of
logarithmic Van Hove singularities (VHS) in the electron density of states.
The scaling
trajectories are presented for different magnetic phases.
It is demonstrated that VHS lead to a considerable increase
of the non-Fermi-liquid behavior region owing to
softening of magnon branches during the renormalization process.
 Although the
effective coupling constant remains moderate, renormalized magnetic moment
and spin-fluctuation frequency can decrease by several orders of magnitude.
A possible
application to $f$-systems and weak itinerant magnets is discussed.
\end{abstract}

\maketitle

%\address{Institute of Metal Physics, 620219 Ekaterinburg, Russia}

\section{Introduction}

Anomalous rare-earth and sctinide compounds are studied extensively starting
from the middle of 1980s \cite{Stewart,Brandt}. They include so-called Kondo
lattices (with moderately enhanced electronic specific heat) and
heavy-fermion systems demonstrating a huge linear specific heat. Main role
in the physics of the Kondo lattices \cite{Brandt,Zwicknagl,Col2}
belongs to the interplay of the on-site
Kondo screening and intersite exchange interactions. Following to Doniach
criterion \cite{Don}, it was believed in early works \cite{Brandt}
that the total suppression of
either magnetic moments or the Kondo anomalies takes place. However, later
experimental data and theoretical investigations made clear that the
Kondo lattices as a rule demonstrate magnetic ordering or are close to this.
This concept was consistently formulated and justified in a series of the
papers \cite{IKFTT,IKZ1,Von2,kondo} treating the mutual
renormalization of two characteristic energy scales: the Kondo temperature $%
T_{K}$ and spin-fluctuation frequency $\overline{\omega }$. A simple scaling
consideration of this renormalization process in the $s-f$ exchange model
\cite{kondo,kondo1} yields, depending on the values of bare
parameters, both the \textquotedblleft usual\textquotedblright\ states (a
non-magnetic Kondo lattice or a magnetic state with weak Kondo corrections) and
the peculiar magnetic Kondo-lattice state. In the latter regime, small
variations of parameters result in strong changes of the ground-state
moment. Thereby high sensitivity of the ground-state moment to external factors
like pressure and doping by a small amount of impurities (a characteristic
feature of heavy fermion magnets) is naturally explained.

During 1990s, a number of anomalous $f$-systems (U$_{x}$Y$_{1-x}$Pd$_{3}$,
UPt$_{3-x}$Pd$_{x}$, UCu$_{5-x}$Pd$_{x}$, CeCu$_{6-x}$Au$_{x}$, U$_{x}$Th$%
_{1-x}$Be$_{13},$CeCu$_{2}$Si$_{2},$CeNi$_{2}$Ge$_{2},$ Ce$_{7}$Ni$_{3}$
etc.) demonstrating the non-Fermi-liquid (NFL) behavior have become a
subject of great interest (see, e.g., the reviews \cite{Maple,Stewart1}).
These systems possess unusual logarithmic or power-law temperature
dependences of electronic and magnetic properties.

Various mechanisms were proposed to describe  the NFL behavior \cite{Proc},
including two-channel Kondo scattering \cite{Tsv,Col}, ``Griffiths
singularities'' in disordered magnets \cite{Castr}, strong spin
fluctuations near a quantum magnetic phase transition \cite{Col1,Vojta2}.

It is important that experimentally the NFL behavior (as well as heavy-fermion
behavior) is
typical for systems lying on the boundary of magnetic ordering and
demonstrating strong spin fluctuations \cite{Maple,Col2}.

The  NFL behavior close to the quantum phase transition was theoretically
studied  in a number of works \cite{Kim,Si,Vojta,Vojta2}.
In particular, renormalization group investigations near the quantum phase transitions
connected with the topology of the Fermi surface were performed \cite{Si}.
A scaling consideration of the Kondo lattices with account of
singularities in the spin excitation spectral function (which are owing to
Van Hove singularities in the magnon spectrum) yields the NFL behavior in an
extremely narrow interval of bare parameters only \cite{kondo}. As
demonstrated in Ref. \cite{nfl}, when taking into account renormalization of
spin-excitation damping, the region can become considerably broader.

The systems under consideration demonstrate both local-moment and itinerant
features.
Moreover, large linear specific heat and NFL behavior is observed
also in some $d$-systems including layered ruthenates Sr$_{2}$RuO$_{4}$ \cite%
{ruthenates} and Sr$_{3}$Ru$_{2}$O$_{7}$ \cite{ruthenates1}.

It is well known that magnetism of itinerant systems is intimately related
to the presence of Van Hove singularities (VHS) near the Fermi level.
Therefore, it is instructive to treat the Kondo effect in systems with a
singular electron spectrum. This is the aim of the present paper.

It is evident that the Kondo effect in such systems has a number of peculiar
features. In particular, for the logarithmically divergent density of states%
\begin{equation*}
\rho (E)=A\ln \frac{D}{B|E|}
\end{equation*}%
(the energy is referred to the Fermi level, $D$ is the half-bandwidth, the
constants $A$ and $B$ are determined by the band spectrum) the Kondo
singularities at $E_{F}$ become double-logarithmic. Perturbation expansion
yields a non-standard expression for the one-centre Kondo temperature,
\begin{equation}
T_{K}\varpropto D\exp \left[ -1/(AI)^{1/2}\right] ,  \label{perturb1}
\end{equation}%
instead of the result for the smooth density of states, $T_{K}\varpropto
D\exp [1/2I\rho (0)]$, $I$ being the $s-f$ exchange parameter. The
logarithmic divergence in $\rho (E)$ is typical for the two-dimensional case (in
particular, for the layered ruthenates). However, similar strong
Van Hove singularities can occur also in some three-dimensional systems like
Pd alloys and weak itinerant ferromagnets ZrZn$_{2}$ and TiBr$_{2}$ \cite%
{VKT,pickett}.

In the present work we consider the Kondo problem and the NFL behavior with
the singular electron density of states for the lattice of $d(f)$-spins where a
competition with spin dynamics takes place.

In Sect. 2 the renormalization group equations in the presence of VHS are
presented. In Sect.3  the scaling behavior for a paramagnet and
for magnetic phases with account of spin-excitation damping is considered.
In Sect. 4
we treat the scaling behavior for the magnetic phases with account of the
incoherent contribution to spin spectral function. In Conclusions,
various electron properties and general physical picture of
magnetism are discussed.

\section{The scaling equations in the presence of Van Hove singularities}

We use the $s-d(f)$ exchange model of a Kondo lattice
\begin{equation}
H=\sum_{\mathbf{k}\sigma }t_{\mathbf{k}}c_{\mathbf{k}\sigma }^{\dagger }c_{%
\mathbf{k}\sigma }^{{}}-I\sum_{i\alpha \beta }\mathbf{S}_{i}%
\mbox {\boldmath
$\sigma $}_{\alpha \beta }c_{i\alpha }^{\dagger }c_{i\beta }^{{}}+\sum_{%
\mathbf{q}}J_{\mathbf{q}}\mathbf{S}_{\mathbf{-q}}\mathbf{S}_{\mathbf{q}}
\end{equation}%
where $t_{\mathbf{k}}$ is the band energy, $\mathbf{S}_{i}$ and $\mathbf{S}_{%
\mathbf{q}}$ are spin-density operators and their Fourier transforms, $J_{%
\mathbf{q}}$ are the intersite exchange parameters, $\sigma $ are the Pauli
matrices.

The density of states corresponding to the spectrum $t_{\mathbf{k}}$ is
supposed to contain a Van Hove singularity near the Fermi level. In
particular, for the square lattice with the spectrum
\begin{equation*}
t_{\mathbf{k}}=2t(\cos k_{x}+\cos k_{y})+4t^{\prime }(\cos k_{x}\cos k_{y}+1)
\end{equation*}%
we have the density of states%
\begin{equation}
\rho (E)=\frac{1}{2\pi ^{2}\sqrt{t^{2}+Et^{\prime }-4t^{\prime 2}}}K\left(
\sqrt{\frac{t^{2}-(E-8t^{\prime })^{2}/16}{t^{2}+Et^{\prime }-4t^{\prime 2}}}%
\right) \simeq \frac{1}{2\pi ^{2}\sqrt{t^{2}-4t^{\prime 2}}}\ln \frac{16%
\sqrt{t^{2}-4t^{\prime 2}}}{|E|}
\end{equation}%
where $K(E)$ is the complete elliptic integral of the first kind, the
bandwidth is determined by $|E-8t^{\prime }|<4|t|$. For $t^{\prime }=0$ we
derive%
\begin{equation}
\rho (E)=\frac{2}{\pi ^{2}D}K\left( \sqrt{1-\frac{E^{2}}{D^{2}}}\right)
\simeq \frac{2}{\pi ^{2}D}\ln \frac{4D}{|E|},~|E|<D=4|t|
\end{equation}%
so that, according to (\ref{perturb1}),
\begin{equation}
T_{K}\varpropto D\exp \left[ -\left( \frac{\pi ^{2}D}{2I}\right) ^{1/2}%
\right] .  \label{perturb}
\end{equation}%
Note that the expression for the Kondo temperature in the parquet
approximation has a different form \cite{gogolin}.
\begin{equation}
T_{K}\varpropto D\exp \left[ -\frac{1}{(AI/2)^{1/2}}\right] =D\exp \left[
-\left( \frac{\pi ^{2}D}{I}\right) ^{1/2}\right]  \label{gogolin}
\end{equation}%
However, the expression (\ref{perturb}) agrees with the numerical Wilson
renormalization group calculation for the square lattice \cite{zhuravlev},
unlike the result (\ref{gogolin}); the corresponding problems of the parquet
approximation in the Hubbard model are discussed in the works \cite{parquet}.

In Refs.\cite{kondo,kondo1,nfl}, the interplay of the Kondo effect and
intersite interactions was investigated by the renormalization group method.
This starts from the second-order perturbation theory with the use of the
equation-of-motion method (within the diagram technique for pseudofermions,
such an approximation corresponds to the one-loop scaling).

We apply the \textquotedblleft poor man scaling\textquotedblright\ approach
\cite{And}. This considers the dependence of effective (renormalized) model
parameters on the cutoff parameter $C$ which occurs at picking out the
singular contributions from the Kondo corrections to the effective coupling
and spin-fluctuation frequencies.
Using the results of Refs.\cite{kondo,nfl} we can write down
the system of scaling equations in the case of the Kondo lattice for various
magnetic phases.

In the calculations below we use the density of states for a square lattice
with $t^{\prime }=0$ both in the two-dimensional (2D) and three-dimensional
(3D) cases as a phenomenological one, so that%
\begin{equation}
\varrho (E)=\varrho (-D)F(E),\varrho (-D)=\frac{2\ln 4}{\pi ^{2}D},\
F(E)=\ln \frac{D}{|E|}+1
\end{equation}

We adopt the definition of the effective (renormalized) and bare $s-f$
coupling constant%
\begin{equation}
g_{ef}(C)=-2\varrho I_{ef}(C),\ g=-2I\varrho ,~\rho =\varrho (-D)
\end{equation}%
where $C\rightarrow -0$ is a flow cutoff parameter. Other relevant variables
are the characteristic spin-fluctuation energy $\overline{\omega }_{ef}(C)$ and
magnetic moment $\overline{S}_{ef}(C)$.

To find the equation for $I_{ef}(C)$ we have to treat the electron
self-energy. For a ferromagnet (the case of an antiferromagnet is considered
in a similar way, see Ref.\cite{kondo}) the second-order Kondo contribution
reads
\begin{equation}
\Sigma _{\mathbf{k\pm }}^{(2)}(E)=\pm 2I^{2}\overline{S}\sum_{\mathbf{q}}%
\frac{n_{\mathbf{k-q}}}{E-t_{\mathbf{k-q}}\pm \omega _{\mathbf{q}}}.
\end{equation}%
where $n_{\mathbf{k}}=f(t_{\mathbf{k}})$ is the Fermi function. Then we have
\begin{equation}
\delta I_{ef}=[\Sigma _{\mathbf{k\downarrow }}^{(2)}(E)-\Sigma _{\mathbf{%
k\uparrow }}^{(2)}(E)]/(2S)
\end{equation}
Picking out
in the sums the contribution of intermediate electron states near the Fermi
level with $C<t_{\mathbf{k+q}}<C+\delta C$ we obtain
\begin{equation}
\delta I_{ef}(C)=2\rho F(C)I^{2}\eta (-\frac{\overline{\omega }}{C})\delta
C/C  \label{ief}
\end{equation}%
where $\eta (x)$ is a scaling function which satisfies the condition $\eta
(0)=1$ which guarantees the correct one-impurity limit.
In the magnetically ordered
phase, $\overline{\omega }$ is the magnon frequency $\omega _{\mathbf{q}%
},$ which is averaged over the wavevectors $\mathbf{q=}2\mathbf{k}$
where $\mathbf{k}$ runs over the Fermi surface (for simplicity we use a
spherical Fermi surface). In the paramagnetic  phase (the problem of
localized moment screening) $\overline{\omega }$ is determined from the
second moment of the spin spectral density.

Now we treat the singular correction to $\overline{\omega }_{ef}$ and the
effective magnetic moment $\overline{S}_{ef}$. We have within the spin-wave
picture
\begin{equation}
\delta \bar{S}=-\sum_{\mathbf{q}}\delta \langle b_{\mathbf{q}}^{\dagger }b_{%
\mathbf{q}}\rangle   \label{Ssf}
\end{equation}
The singular contribution to magnon occupation numbers occurs owing to the
electron-magnon interaction. Calculation  for a ferromagnet from the
corresponding magnon Green's function yields \cite{kondo}
\begin{equation}
\delta \langle b_{\mathbf{q}}^{\dagger }b_{\mathbf{q}}\rangle =I^{2}S\sum_{%
\mathbf{k}}\frac{n_{\mathbf{k}}(1-n_{\mathbf{k-q}})}{(t_{\mathbf{k}}-t_{%
\mathbf{k-q}}-\omega _{\mathbf{q}})^{2}}  \label{bbf}
\end{equation}

We see that, when considering characteristics of localized-spin subsystem,
the lowest-order Kondo corrections originate from double integrals over both
electron and hole states. Then we have to introduce two cutoff parameters $%
C_{e}$ and $C_{h}$ with $C_{e}+C_{h}=C$ ($C$ is the cutoff parameter for the
electron-hole excitations), $\delta C_{e}=-\delta C_{h}~$to obtain
\begin{equation}
\delta \overline{S}_{ef}(C)/S=2\rho ^{2}I^{2}F(C/2)F(-C/2)\eta (-\frac{%
\overline{\omega }}{C})\delta C/C  \label{sef}
\end{equation}%

The renormalization of spin-wave frequency owing to magnon-magnon scattering
is given by%
\begin{equation}
\delta \omega _{\mathbf{q}}/\omega _{\mathbf{q}}=-a_{\mathbf{q}}\delta \langle b_{\mathbf{%
q}}^{\dagger }b_{\mathbf{q}}\rangle /S
\end{equation}%
Further on we pass to the magnon frequency averaged over the Fermi surface.
Then we have ($a_{\mathbf{q}}\rightarrow a$)%
\begin{equation}
\delta \overline{\omega }_{ef}(C)/\overline{\omega }=a\delta \overline{S}%
_{ef}(C)/S=2a\rho ^{2}I^{2}F(C/2)F(-C/2)\eta (-\frac{\overline{\omega }}{C}%
)\delta C/C  \label{wef}
\end{equation}%
The latter result holds for all magnetic phases with $a=1-\alpha $ for the
paramagnetic (PM) phase, $a=1-\alpha ^{\prime }$ for the antiferromagnetic
(AFM) phase, $a=2(1-\alpha ^{\prime \prime })$ for the ferromagnetic (FM)
phase. Here $\alpha ,\alpha ^{\prime },\alpha ^{\prime \prime }$ are some
averages over the Fermi surface (see Ref.\cite{kondo}), $\alpha ^{\prime }=0$
in the nearest-neighbor approximation.
This approximation enables us to use a single renormalization parameter,
rather than the whole function of ${\bf q}$.
For simplicity, we put in numerical
calculations below $a=1$ (although the deviation $1-a$ just determines
critical exponents for physical properties, see Ref.\cite{nfl} and
Conclusions).

The scaling picture (which determines the NFL behavior)
is influenced by not only real, but also by  imaginary part
of the spin-fluctuation energy.
The latter is even  dominating in the paramagnetic phase (e.g., in the Heisenberg model
a spin-diffusion picture can be adopted at high temperatures).
In the  magnetically ordered phases, the damping comes from paramagnon-like excitations.
They can be taken into account starting from the magnon picture of
the localized-spin excitation spectrum.
In the $s-f$ exchange model the damping is proportional
to $I^{2}$ and to the magnon frequency (for the details, see Ref.\cite{nfl}).
The dependence of the damping on the magnetic moment $\overline{S}$
(which is strongly renormalized) is crucial for the size of the NFL region.
The calculation of the damping  in the second
order in $I$ yields the contributions of the order of both $I^{2}\overline{S}$
and $I^{2}$ \cite{Aus,afm} (formally, they correspond to the first and
second order in the quasiclassical parameter $1/2S$). The corresponding
problems from a semiphemenological point of view are discussed in Ref.\cite%
{sokol}. Similar to Ref.\cite{nfl},  here we do not
introduce the additional factor of $\overline{S}$ to obtain in terms of
renormalized quantities%
\begin{equation}
\overline{\gamma }_{ef}(C)=kF(C/2)F(-C/2)g_{ef}^{2}(C)\overline{\omega }%
_{ef}(C),
\end{equation}%
the factor $k$ being determined by the bandstructure and magnetic ordering;
we put in numerical calculations $k=0.5$.

When taking into account spin-wave damping $\overline{\gamma }$ we have
\begin{equation*}
\eta \left( \frac{\overline{\omega }_{ef}(C)}{|C|}\right) \rightarrow \eta
\left( \frac{\overline{\omega }_{ef}(C)}{|C|},\frac{\overline{\gamma }%
_{ef}(C)}{|C|}\right)
\end{equation*}%
Replacing in the Kondo corrections $g\rightarrow g_{ef}(C),\,\overline{%
\omega }\rightarrow \overline{\omega }_{ef}(C)$ we derive the set of scaling
equations with account of the energy dependence of the electron density of
states:
\begin{eqnarray}
\partial g_{ef}(C)/\partial C &=&F(C)\Lambda ,  \label{gl} \\
\partial \ln \overline{\omega }_{ef}(C)/\partial C &=&-aF(C/2)F(-C/2)\Lambda
/2,  \label{sl} \\
\partial \ln \overline{S}_{ef}(C)/\partial C &=&-F(C/2)F(-C/2)\Lambda /2
\label{ssl}
\end{eqnarray}%
with
\begin{equation*}
\Lambda =\Lambda (C,\overline{\omega }_{ef}(C),\overline{\gamma }_{ef}(C))=%
\frac{g_{ef}^{2}(C)}{|C|}\eta \left( \frac{\overline{\omega }_{ef}(C)}{|C|},%
\frac{\overline{\gamma }_{ef}(C)}{|C|}\right) .
\end{equation*}%

Similar equations can be obtained for the general $SU(N)$ Coqblin-Schrieffer
model \cite{kondo} where $a/2\rightarrow a/N$ (there are some peculiarities
for FM case owing to asymmetry of spin-up and spin-down states).

One can see that the renormalizations of the spin-fluctuation energy $%
\overline{\omega }_{ex}(C)$ and the damping are more strong than that of  $%
g_{ef}(C)$\ owing to the factors of $F(\pm C/2).$ We obtain from (\ref{sl}), (%
\ref{ssl})
\begin{equation}
\frac{\overline{S}_{ef}(C)}{S}=\left( \frac{\overline{\omega }_{ef}(C)}{%
\overline{\omega }}\right) ^{1/a}  \label{sscc}
\end{equation}%
However, the simple expression for $\rho =~$const,
\begin{equation}
\overline{\omega }_{ef}(C)=\overline{\omega }\exp (-a[g_{ef}(C)-g]/2),
\label{linsmooth}
\end{equation}%
does not hold for the logarithmic density of states; no simple relation with
the quantity $\widetilde{g}_{ef}(C)=F(C)g_{ef}(C)$ is  obtained either.
Expanding in $1/\ln |D/C|,$ we derive%
\begin{equation}
\overline{\omega }_{ef}(C)\simeq \overline{\omega }\exp \left( -\frac{a}{2}%
\int_{-D}^{C}\frac{dC^{\prime }}{C^{\prime }}[g_{ef}(C)-g_{ef}(C^{\prime
})]-a\ln 2[g_{ef}(C)-g]\right)  \label{compl}
\end{equation}

Now we treat the scaling functions $\eta $. In the paramagnetic case we use
the spin-diffusion approximation (dissipative spin dynamics) to obtain (cf.
Ref.\cite{kondo})%
\begin{equation}
\eta ^{PM}(\frac{\overline{\omega }}{C})=\left\langle \frac{1}{1+\mathcal{D}(%
\mathbf{k-k}^{\prime })^{2}/C^{2}}\right\rangle _{t_{k}=t_{k^{\prime }}=0},~%
\overline{\omega }=4\mathcal{D}k_{F}^{2}
\end{equation}%
where $\mathcal{D}$ is the spin diffusion constant, the averages go over the
Fermi surface. Integration yields%
\begin{equation*}
\eta ^{PM}(x)=\left\{
\begin{array}{cc}
\arctan x/x & d=3 \\
\{\frac{1}{2}[1+(1+x^{2})^{1/2}]/(1+x^{2})\}^{1/2} & d=2%
\end{array}%
\right.
\end{equation*}%
In the FM and AFM phases for simple magnetic structures we have
\begin{equation}
\eta \left( \overline{\omega }_{ef}/|C|,\overline{\gamma }_{ef}/|C|\right) =%
{\rm Re}\left\langle \left( 1-(\omega _{\mathbf{k-k}^{\prime
}}^{{}}+i\gamma _{\mathbf{k-k}^{\prime }}^{{}})^{2}/C^{2}\right)
^{-1}\right\rangle _{t_{k}=t_{k^{\prime }}=0}  \label{etafm}
\end{equation}%
For an isotropic 3D ferromagnet integration in (\ref%
{etafm}) for $\gamma =$ const and quadratic spin-wave spectrum $\omega _{%
\mathbf{q}}\propto q^{2}$ yields
\begin{equation}
\eta ^{FM}(x,z)=\frac{1}{4x}\ln \frac{(1+x)^{2}+z^{2}}{(1-x)^{2}+z^{2}}
\label{intfm}
\end{equation}%
where $x=\overline{\omega }_{ef}/|C|,z=$ $\overline{\gamma }_{ef}/|C|.$
Although details of the spin-wave spectrum are not reproduced in such an approach,
the renormalization of spin-wave frequency (which is an cutoff for the Kondo divergences)
is adequately reproduced.

In the 2D case we obtain in the same approximation
\begin{equation}
\eta ^{FM}(x,z)=\frac{1}{2}{\rm Re}%
\{[(1+iz)(1+iz-x)]^{-1/2}+[(1-iz)(1-iz+x)]^{-1/2}\}
\end{equation}%
For an antiferromagnet integration in (\ref{etafm}) with the linear spin-wave
spectrum $\omega _{\mathbf{q}}\propto cq$ gives
\begin{equation}
\eta ^{AFM}(x,z)=\left\{
\begin{array}{cc}
-\frac{1}{2x^{2}}\text{ }\{\ln [(1+z^{2}+x^{2})^{2}-4x^{2}]-2\ln (1+z^{2})\}
& d=3 \\
2{\rm Im}(x^{2}-1-2iz+z^{2})^{-1/2} & d=2%
\end{array}%
\right.  \label{intafm}
\end{equation}%
which modifies somewhat the results of Ref.\cite{kondo}.

\begin{figure}[tbp]
\includegraphics[width=0.45\columnwidth]{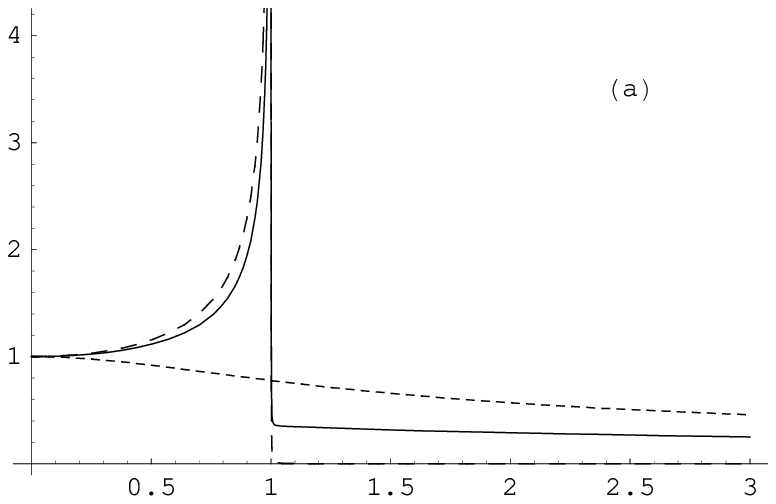}
\includegraphics[width=0.45\columnwidth]{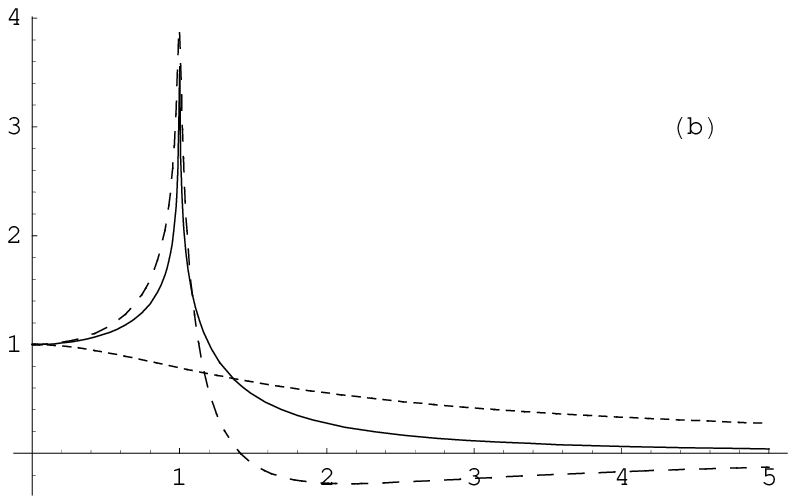}
\caption{
The scaling functions $\eta (x)$ for a ferromagnet (solid line),
antiferromagnet (long-dashed line) and paramagnet (short-dashed line) in the
2D (a) and 3D (b) cases
}
\label{fig:1}
\end{figure}

The plot of the functions $\eta (x)=$ $\eta (x,0)$ for different magnetic
phases is shown in Fig.1. Note that for a 2D antiferromagnet $\eta (x)$
vanishes discontinuosly at $x>1$. However, a smooth non-zero contribution
can occur for more realistic models of magnon spectrum.

\section{The scaling behavior in paramagnetic and magnetic phases}

We start the discussion of scaling behavior
from the simple case of the Coqblin-Schrieffer model in the limit $%
N\rightarrow \infty .$ Then the renormalization of the magnon frequency is
absent and the scaling behavior can be investigated analytically, similar to
the \cite{kondo}. We have
\begin{equation}
1/g_{ef}(C)-1/g=G(C)=-\int_{-D}^{C}\frac{dC^{\prime }}{C^{\prime }}%
F(C^{\prime })\eta (-\frac{\overline{\omega }}{C^{\prime }})  \label{intG}
\end{equation}%
The equation (\ref{intG}) can be used even for $N=2$ provided that $g$ is
considerably smaller than the critical value $g_{c}$. The effective coupling
 $g_{ef}(C)$ begins to deviate strongly from its one-impurity
behavior
\begin{equation}
1/g_{ef}(C) \simeq 1/g-\frac{1}{2}\ln ^{2}|D/C|
\end{equation}%
at $|C|\sim \overline{\omega }.$  The boundary of the strong coupling
region (the renormalized Kondo temperature) is determined by $%
G(C=-T_{K}^{\ast })=-1/g.$ Of course, $T_{K}^{\ast }$ means here only some
characteristic energy scale extrapolated from high temperatures, and the
detailed description of the ground state requires a more detailed
consideration. In the PM, FM and 2D AFM phases spin dynamics suppresses $%
T_{K}^{\ast }.$ To leading order in $\ln (D/\overline{\omega })$ we have%
\begin{equation*}
T_{K}^{\ast }\simeq (T_{K}^{2}-\overline{\omega }^{2})^{1/2}
\end{equation*}%
with $T_{K}$ given by (\ref{perturb}). [However, owing to the minimum of the
scaling function (\ref{intafm}) (Fig. 1b), in the 3D AFM case spin dynamics
at not large $\overline{\omega }$ results in and increase of $T_{K}^{\ast }$%
.]

Provided that the strong coupling regime does not occur, i.e. $g$ is smaller
than the critical value $g_{c}$, $g_{ef}(C\rightarrow 0)$ tends to a finite
value $g^{\ast }.$ To leading order in $\ln (D/\overline{\omega })$ we have
\begin{equation}
1/g_{c}=\frac{1}{2}\ln ^{2}\frac{D}{\overline{\omega }}=\frac{1}{2}\lambda
^{2}
\end{equation}%
(which yields also a rough estimate of $g_{c}$ for $N=2$). An account of
next-order terms results in an appreciable dependence on the type of
magnetic ordering and space dimensionality. For PM, FM and 2D AFM phases the
critical value $g_{c}$ is given by $1/g_{c}=-G(0)$, and in the 3D AFM case $%
g_{c}$ is determined by the minimum of the function $G(C)$.

\begin{figure}[tbp]
\includegraphics[width=0.45\columnwidth]{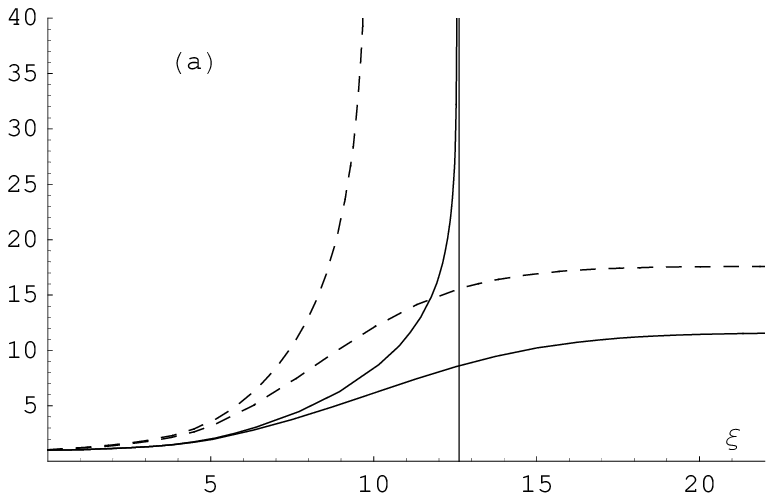}
\includegraphics[width=0.45\columnwidth]{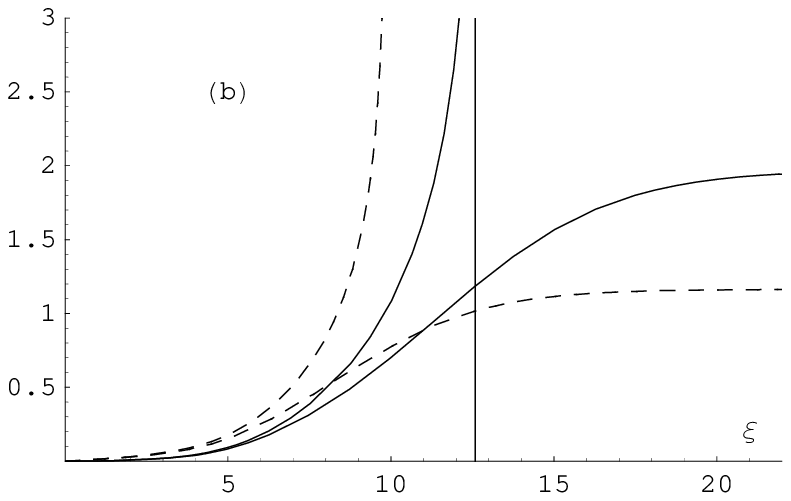}
\caption{
The scaling trajectories $g_{ef}(\xi =\ln |D/C|)/g$ (a)  and the
corresponding dependences $\ln (S/\overline{S}(\xi ))=\ln (\overline{\omega }%
/\overline{\omega }_{ef}(\xi ))$ (b) for 2D paramagnets with the logarithmic
density of states (solid lines, $g=0.029<g_{c},g=0.030>g_{c}$) and constant
density of states (dashed lines, $g=0.14<g_{c},g=0.145>g_{c}$),
$a=1,\lambda =\ln (D/\overline{\omega })=5$
}
\label{fig:2}
\end{figure}

The numerical calculations for $N=2$ were performed for $\lambda =\ln (D/%
\overline{\omega })=5.$ The plots are presented in Figs.2-5
for both smooth and singular bare
densities of states. We compare these cases in the
same relative interval of the coupling constant $|g-g_{c}|/g_{c}$. The
scaling process for finite $N$  in the former case is
described in Ref. \cite{kondo}. It turns out that the qualitative
picture near the
critical value of the coupling constant is rather universal for the same
interval $|g-g_{c}|/g_{c}$ depending mainly on the scaling function and the
damping, but not on the details of the bare electron density of states (even
on its singularities). The shift of the Van Hove singularity below the Fermi
level does not influence strongly the results too, although we cross the
singularity during the scaling process (the singularity is in fact
integrable).

An important quantitative difference in the presence of VHS is that the
renormalized coupling constant is considerably smaller. Moreover, relative
renormalization of the coupling constant is also smaller (Fig.2-4). This
makes using perturbation theory and lowest-order scaling analysis
more physically
reliable than for the smooth density of states.

At the same time, in the presence of VHS the renormalization of
spin-fluctuation frequency (and of magnetic moment) becomes larger (a
\textquotedblleft soft mode\textquotedblright\ situation, which favors a NFL
behavior). Formally, this is due to that the derivative $\partial \ln
\overline{\omega }_{ef}(C)/\partial C$ is proportional to $F^{2}(C/2)$, and $%
\partial g_{ef}(C)/\partial C$ to $F(C)$ only [see Eqs. (\ref{sl})-(\ref%
{ssl})]. Such a situation is similar to the scaling in the large-$l$ limit ($%
l$ is the number of scattering channels for conduction electrons) where the
effective coupling is not renormalized, $(2l+1)g^{2}/2=\widetilde{g}^{2}=~$%
const, since $a\rightarrow a(2l+1)$ \cite{kondo}.

For a paramagnet with pure dissipative dynamics, the one-impurity behavior $%
1/g_{ef}(\xi =\ln |D/C|)=1/g-\xi $ is changed at $\xi \simeq \lambda $ by a
NFL-like (smeared quasi-linear) region where
\begin{equation}
~\ln [\overline{\omega }/\overline{\omega }_{ef}(\xi )]=a\ln [S/\overline{S}%
(\xi )]\simeq (aA/2)\xi  \label{lina}
\end{equation}%
with $A<2/a$. Such a behavior takes place both for $g<g_{c}$ and $g>g_{c}$
in a wide region of $\xi ,$ i.e. up to rather low temperatures (Fig. 2b, for
a discussion of physical properties see Conclusions).

\begin{figure}[tbp]
\includegraphics[width=0.45\columnwidth]{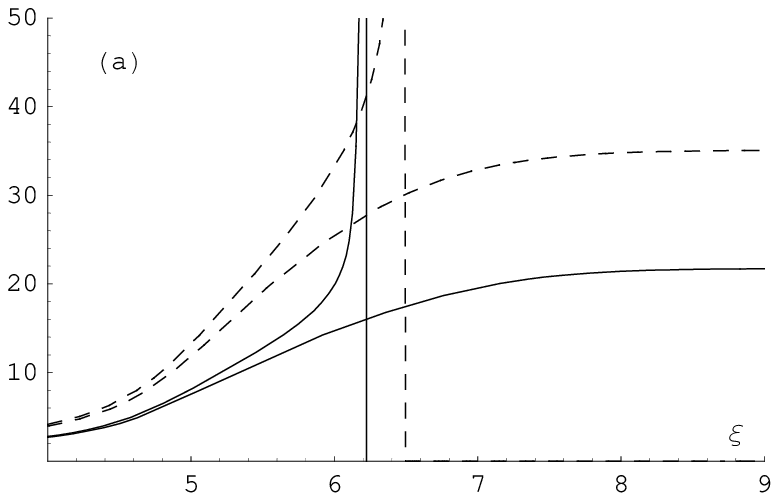}
\includegraphics[width=0.45\columnwidth]{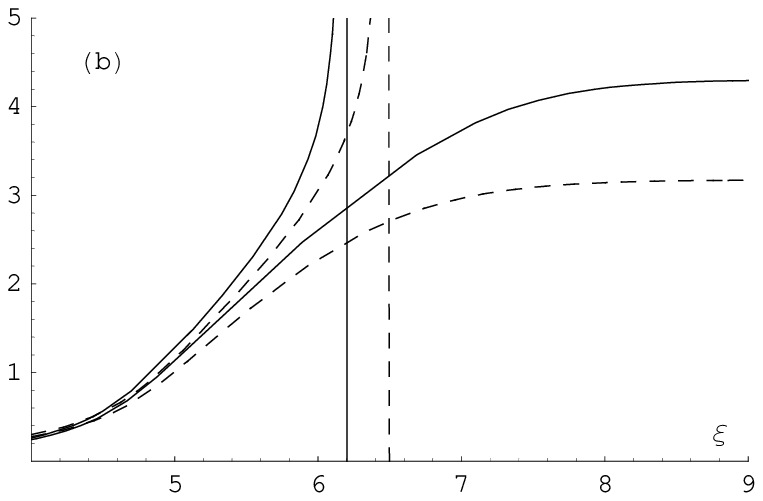}
\caption{
The scaling trajectories $g_{ef}(\xi )/g$ (a) and the corresponding
dependences $\ln (S/\overline{S}(\xi ))=\ln (\overline{\omega }/\overline{%
\omega }_{ef}(\xi ))$ (b) for 2D ferromagnets with the logarithmic
(solid lines, $g=0.052<g_{c},g=0.053>g_{c}$) and constant
density of states (dashed lines, $g=0.186<g_{c},g=0.189>g_{c}$), $k=0.5$,
other parameters being the same as in Fig.2
}
\label{fig:3}
\end{figure}

\begin{figure}[tbp]
\includegraphics[width=0.45\columnwidth]{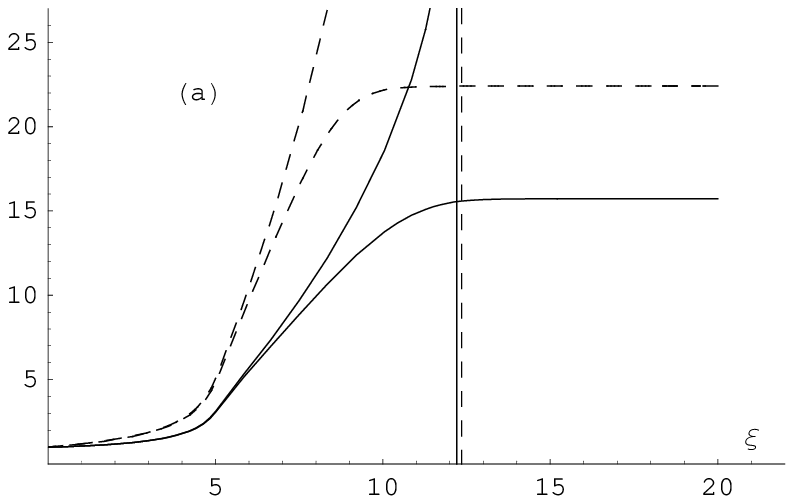}
\includegraphics[width=0.45\columnwidth]{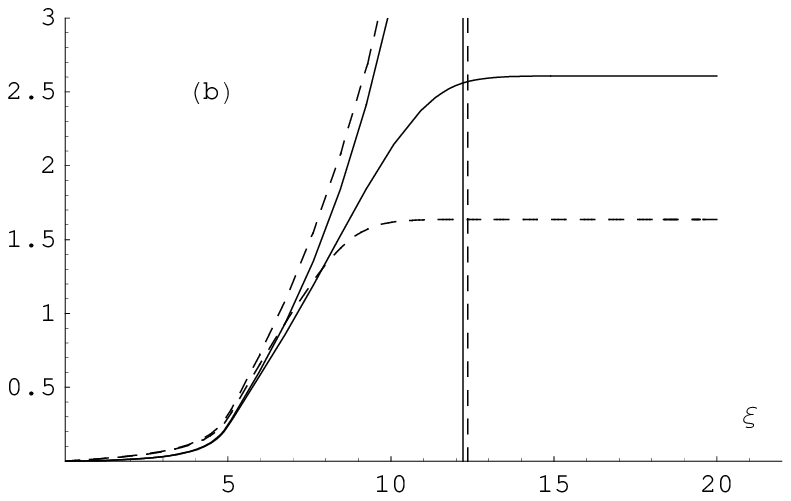}
\caption{
The scaling trajectories $g_{ef}(\xi )/g$ (a) and
the corresponding dependences $\ln (S/\overline{S}%
(\xi ))=\ln (\overline{\omega }/\overline{\omega }_{ef}(\xi ))$ (b)
for 2D antiferromagnets
with the logarithmic  (solid lines, $%
g=0.0361<g_{c},g=0.0365>g_{c}$) and constant density of states (dashed
lines, $g=0.153<g_{c},g=0.155>g_{c}$)
}
\label{fig:4}
\end{figure}

For magnetic phases, the singularities of the scaling function $\eta
(x\rightarrow 1)$ play the crucial role. A rather distinct NFL behavior
takes place in a more narrow region where the argument of the function $\eta
$ is fixed at the singularity during the scaling process, so that
\begin{equation}
\overline{\omega }_{ef}(C)\simeq |C|~,~\ln [\overline{\omega }/\overline{%
\omega }_{ef}(\xi )]\simeq \xi .  \label{w=c}
\end{equation}%
Then for a smooth density of states we obtain from (\ref{linsmooth})
\begin{equation}
g_{ef}(\xi )-g\simeq 2(\xi -\lambda )/a  \label{linnn}
\end{equation}%
However, in the presence of VHS the relation (\ref{compl}) between $%
\overline{\omega }_{ef}(C)$ and $g_{ef}(C)$ is more complicated.

In the case of a constant magnon damping considered in Ref. \cite{kondo},
the region (\ref{w=c}) is not too narrow only provided that the bare
coupling constant $g$ is very close to the critical value $g_{c}$ for the
magnetic instability ($|g-g_{c}|/g_{c}\sim 10^{-4}\div 10^{-6}$). However,
when taking into account the magnon damping renormalization, this region is
considerably wider, although being smeared \cite{nfl}, and the influence of
the energy dependence $\rho (E)$ becomes stronger.

One can see that VHS lead to a considerable increase of the effective moment
$S^{\ast }=\overline{S}(\xi \rightarrow \infty ).$ For the chosen deviation $%
|g-g_{c}|/g_{c}\sim 1\%$ the moment renormalization $S/S^{\ast }$ can make
up one-two orders of magnitude; even for $|g-g_{c}|/g_{c}\sim 10-20\%,~$the
moment can decrease by several times.

Since in 2D antiferromagnet $\eta (x>1)=0,$ a sharp transition to the
saturation plateau occurs, unlike the FM case (cf. Figs.3 and 4).

\begin{figure}[tbp]
\includegraphics[width=0.45\columnwidth]{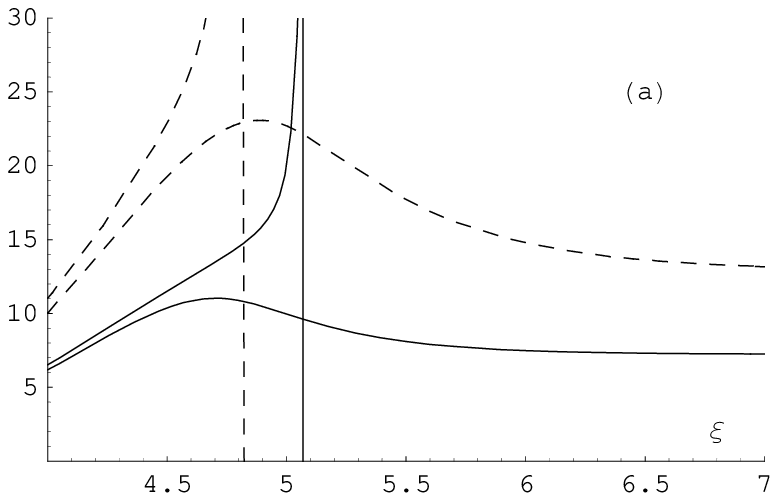}
\includegraphics[width=0.45\columnwidth]{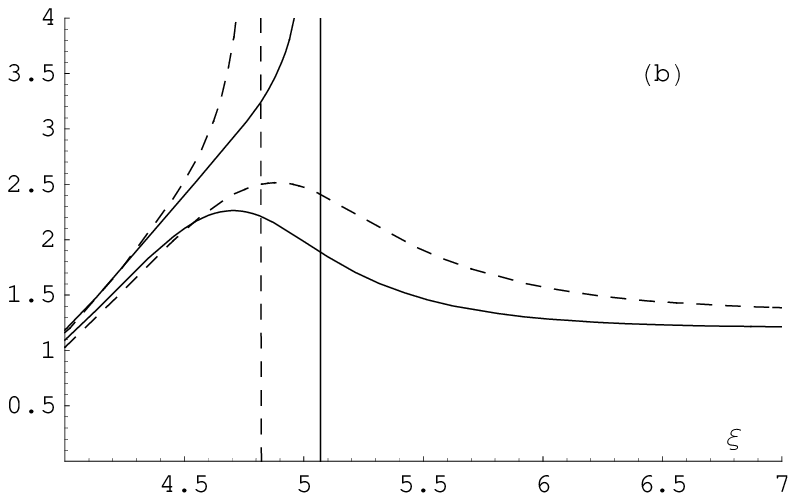}
\caption{
The scaling trajectories $g_{ef}(\xi )/g$ (a) and
the corresponding dependences $\ln (S/\overline{S}(\xi
))=\ln (\overline{\omega }/\overline{\omega }_{ef}(\xi ))$ (b)
for 3D antiferromagnets
with the logarithmic (solid lines, $g=0.072<g_{c},g=0.073>g_{c}$)
and constant density of states (dashed lines,
$g=0.228<g_{c},g=0.231>g_{c}$)
}
\label{fig:5}
\end{figure}

The situation in 3D systems with a logarithmic density of states can be
considered in a similar way. Surprisingly, for 3D ferromagnets to high
accuracy the scaling trajectories (except for a narrow critical region near $%
g_{c}$) and the critical values $g_{c}$ turn out to be very close to 2D case
(for both smooth density of states and with VHS). Therefore we do not show
the corresponding plot. On the other hand, for 3D antiferromagnets VHS does
not lead to suppression of magnon frequencies (Fig.5). This is due to the
influence of the minimum in the scaling function (Fig.1b).

\section{The scaling behavior with account of incoherent contributions}

The magnon approximation used in the previous Section is not quite valid since
this underestimates the role of the damping. In fact, the spin spectral
function should have an intermediate form between large-damping and spin-wave
pictures, containing both coherent (magnon-like) and incoherent
contributions A simple attempt to construct the corresponding scaling
function as a linear combination was performed in Ref. \cite{kondo}. In
particular, in the case of a ferromagnet we have near the magnon pole
\begin{equation}
\langle \langle S_{\mathbf{q}}^{+}|S_{-\mathbf{q}}^{-}\rangle \rangle
_{\omega }=\frac{2\overline{S}\mathcal{Z}_{\mathbf{q}}}{\omega -\omega _{%
\mathbf{q}}^{ef}}+\langle \langle S_{\mathbf{q}}^{+}|S_{-\mathbf{q}%
}^{-}\rangle \rangle _{\omega }^{incoh}  \label{gz}
\end{equation}%
where the inverse residue at the pole is determined by
\begin{equation}
1/\mathcal{Z}_{\mathbf{q}}=1-\left( \frac{\partial \Pi _{\mathbf{q}}(\omega )%
}{\partial \omega }\right) _{\omega =\omega _{\mathbf{q}}}
\end{equation}%
$\Pi _{\mathbf{q}}(\omega )$ being the polarization operator of the magnon
Green's function. Besides that, there exists the singular contribution which
comes from the incoherent (non-pole) part of the spin spectral density. Then
we get
\begin{eqnarray}
\partial g_{ef}(C)/\partial C &=&F(C)\Lambda  \label{gll} \\
\partial \ln \overline{\omega }_{ef}(C)/\partial C &=&-aF(C/2)F(-C/2)\Lambda
/2  \label{wll} \\
\partial (1\emph{/}\mathcal{Z})/\partial C &=&\partial \ln \overline{S}%
_{ef}(C)/\partial C=-F(C/2)F(-C/2)\Lambda /2  \label{zl}
\end{eqnarray}%
where%
\begin{equation}
\Lambda =[g_{ef}^{2}(C)/|C|][\mathcal{Z}\eta _{coh}(\overline{\omega }%
_{ef}(C)/|C|)+(1-\mathcal{Z})\eta _{incoh}(\overline{\omega }_{ef}(C)/|C|)]
\label{l1}
\end{equation}%
with $\eta _{coh}=\eta ^{FM}$. The choice of $\eta _{incoh}$ is a more
difficult problem; here we put simply $\eta _{incoh}=\eta ^{PM}.~$

According to (\ref{zl}) we have
\begin{equation}
\frac{1}{\mathcal{Z}(\xi )}=1+\ln \frac{S}{\overline{S}(\xi )}  \label{1/ZS}
\end{equation}%
Consequently, the increase of magnetic moment owing to the Kondo screening
leads to a considerable logarithmic suppression of magnon contributions to
the spectral density.

The role of the incoherent contribution becomes important only provided that
$S/\overline{S}(\xi )$ and $\mathcal{Z}$ deviate appreciably from unity.
However, such a moment suppression just occurs at passing the region of
singularity in $\eta _{coh}(x=1)$, the further scaling process being
determined by the incoherent contribution.

\begin{figure}[tbp]
\includegraphics[width=0.45\columnwidth]{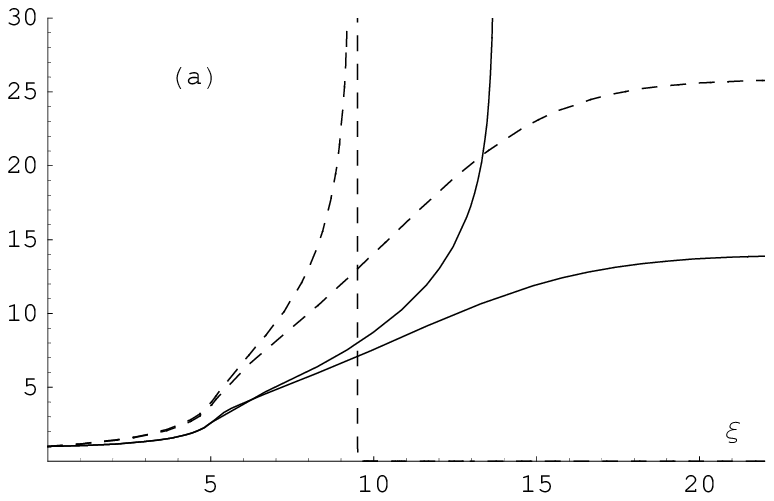}
\includegraphics[width=0.45\columnwidth]{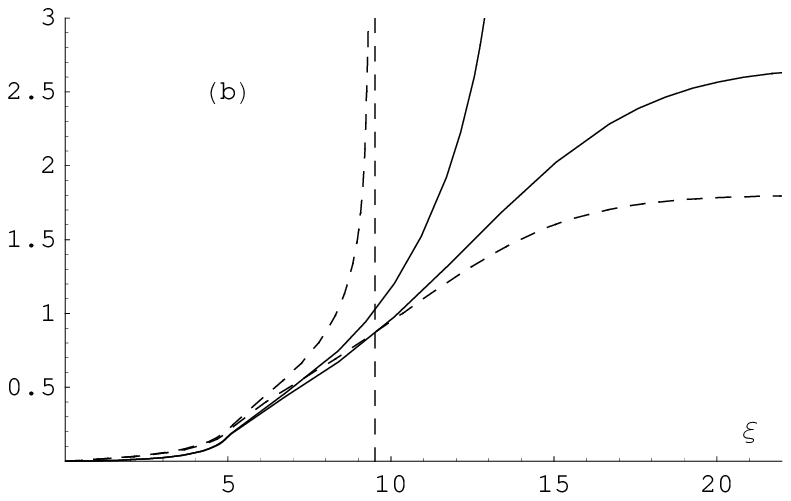}
\caption{
The scaling trajectories $g_{ef}(\xi )/g$ taking into account
incoherent contributions (a) and the corresponding
dependences $\ln (S/\overline{S}(\xi ))=\ln (\overline{\omega }/\overline{%
\omega }_{ef}(\xi ))$ (b)  for 2D ferromagnets with the logarithmic
(solid lines, $g=0.0330<g_{c},g=0.0335>g_{c}$) and constant density
of states (dashed lines, $g=0.143<g_{c},g=0.145>g_{c}$),
$k=0.5,a=1,\lambda =\ln (D/\overline{\omega })=5$
}
\label{fig:6}
\end{figure}

The corresponding scaling trajectories are shown in Fig.6. Now we have a
two-stage renormalization. One can see that the well-linear
\textquotedblleft coherent\textquotedblright\ behavior region (which is
rather narrow in Fig.6) is changed by a PM-like \textquotedblleft
quasi-linear\textquotedblright\ behavior (\ref{lina}) with increasing $\xi.$
This crossover occurs when the function $\eta _{coh}$ reaches its maximum
value at the singularity (cf. Ref.\cite{kondo}).

The \textquotedblleft quasi-linear\textquotedblright\ behavior, although
being somewhat smeared, is pronounced in a considerable region of $\xi $
even for not too small $|g-g_{c}|$. The difference with the PM case is in
that $g_{ef}(\xi )$ increases considerably at the first stage of
renormalization owing to the singularity of the function $\eta _{coh}$.

A similar consideration can be performed for the antiferromagnetic phase
(cf. Ref.\cite{kondo}).
The account of incoherent contribution results in a smearing
of the non-monotonous behavior of $g_{ef}(\xi )$ in the 3D AFM case, so that
at small $|g-g_{c}|$ the maximum in the dependences $g_{ef}(\xi )$ and $\ln [%
\overline{\omega }/\overline{\omega }_{ef}(\xi )]$ vanishes completely.

\section{Discussion and conclusions}

In the general problem of metallic magnetism, the peaks in the bare density
of states (which are usually connected with VHS) near the Fermi level play a
crucial role. Here we have investigated their influence starting from the
Kondo lattice ($s-d(f)$ exchange) model.

For $g\rightarrow g_{c}$ we obtain the magnetic state with small effective
moment $S^{\ast }$ and a NFL-type behavior. The corresponding dependences $%
\overline{S}(T)=\overline{S}_{ef}(|C|\rightarrow T)$ describe an analogue of
the \textquotedblleft temperature-induced magnetism\textquotedblright\ \cite%
{Mor}. Such a picture is based on the many-electron renormalization
(compensation) of localized magnetic moments and differs outwardly from the
ordinary mechanism for weak itinerant ferromagnets with small $\overline{S},$
which are assumed to correspond to the immediate vicinity of the Stoner
instability.

However, the physical difference is not radical. In fact, a continuous
transition exists between the highly-correlated Kondo lattices and the
\textquotedblleft usual\textquotedblright\ itinerant-electron systems. In
particular, one may view Pauli paramagnets as systems with high $T_{K}$ of
order of the Fermi energy; for enhanced Pauli paramagnets like Pd, Pt, UAl$_2$,
where the Curie-Weiss holds at high temperatures, one introduces
instead of the Kondo temperature the so-called spin-fluctuation temperature.
A combined description of the Kondo
lattice state and weak itinerant magnetism has been considered recently by
Ohkawa \cite{Ohkawa}. Remember that the Kondo systems with VHS near the
Fermi level under consideration just possess high values of $T_{K}$ (see the
Introduction).

In this context, it would be instructive to describe weak itinerant magnets
not from the \textquotedblleft band\textquotedblright\ point of view, but
from the side of local magnetic moments which are nearly compensated. Since
a number of cerium NFL systems demonstrate
itinerant-electron behavior \cite{Proc}
and it is customary now to treat UPt$_{3}$, CeSi$_{x}$ and CeRh$_{3}$B$_{2}$
as weak itinerant magnets, the second approach appears already by far less
natural than the first (see Refs.\cite{613,IKZ1}).
From the formal point of view,
perturbation calculations in the Hubbard model, which describes
itinerant-electron systems, are similar to those in the $s-d(f)$ model,
provided that one postulates the existence of local moments. Besides that,
for two-dimensional itinerant systems with strong spin fluctuations the
semiphenomenological spin-fermion model can be used which separates electron
and spin degrees of freedom and is somewhat similar to $s-d$ exchange model
\cite{spin-ferm1,ruthenates2}.

Further on, the question arises about the role which many-electron effects
play in the \textquotedblleft classical\textquotedblright\ weak itinerant
3D magnets like ZrZn${}_{2}$ and TiBe$_{2}$. Indeed, one can hardly believe
that the extremal smallness of $\overline{S}$ in these systems is due to
accidental bare values of $N(E_{F})$ and Stoner parameter. Moreover, the
Stoner criterion is not valid even qualitatively (in particular,  due to
spin fluctuations the critical coupling $U_{c}$ in the Hubbard model is
finite when the Fermi level tends to VHS \cite{ruthenates2}). Thus a scaling
consideration in the presence of VHS would be of interest, especially
with account of chemical potential renormalizations (cf. Ref.\cite{VKT}).
In particular, owing
to VHS the chemical potential can depend weakly on electron concentration (the
pinning phenomenon \cite{pinning}). In the Kondo systems such a treatment
may lead to a renormalization of the scale $|g-g_{c}|$ itself.

Now we discuss some real layered systems. Specific heat is considerably
enhanced in  ruthenates Sr$_{2}$RuO$_{4}.$ A gradual enhancement
of the electronic specific heat and a more drastic increase of the
static magnetic susceptibility were observed in Sr$_{2-y}$La$_{y}$RuO$_{4}$
with increasing $y$. Furthermore, the quasi-2D Fermi-liquid
behavior observed in pure Sr$_{2}$RuO$_{4}$ breaks down near the critical
value $y=0.2$. The enhancement of the density of states can be ascribed to
the elevation of the Fermi energy toward a Van Hove singularity of the
thermodynamically dominant Fermi-surface sheet. The NFL
behavior is attributed to two-dimensional FM fluctuations with
short-range correlations at VHS \cite{ruthenates}.

The bilayered ruthenate system Sr$_{3}$Ru$_{2}$O$_{7}$  in the ground state
is a paramagnetic Fermi liquid with strongly enhanced quasiparticle
masses. The Fermi-liquid region of the phase diagram extends up to 10-15 K
in zero field and is continuously suppressed towards zero temperature upon
approaching the critical field of $B=8$T. In the vicinity of the putative
quantum critical end point, NFL behavior has been observed in
various macroscopic quantities including specific heat, resistivity and
thermal expansion and has been described on the basis of phenomenological
models \cite{ruthenates1}.

We can mention also some layered $f$-systems. The layered Kondo lattice
model was proposed for quantum critical beta-YbAlB$_{4}$ where
two-dimensional boron layers are Kondo coupled via interlayer Yb moments
\cite{YbAlB4}. CeRuPO seems to be one of the rare examples of a ferromagnetic
Kondo lattice where LSDA+U calculations evidence a quasi-2D
electronic band structure, reflecting a strong covalent bonding within the
CeO and RuP layers and a weak ioniclike bonding between the layers \cite%
{CeRuPO}.

To describe layered antiferromagnetic cuprates, the 2D $t-t'$  Hubbard model
is often used which also describes Fermi-liquid and NFL regimes.
Despite  the density-of-state logarithmic singularity, the staggered spin susceptibility
in this model does not diverge within the Fermi-liquid approach,
the reason being the appearance of the logarithmic singularity
in the quasiparticle mass \cite{Hlubina}.
%is somewhat more complicated.
%Unlike ferromagnets, where
%VHS in the electronic spectrum strongly favors magnetic instability,
%antiferromagnetism in a 2D Hubbard model
%does not appear around VHS, the reason being the
%Because of the mass renormalization,
The effective mass renormalization is beyond our lowest-order (one-loop) scaling
consideration, but may play a role at an accurate treatment.
The two-loop considerations  of the flat-Fermi-surface and  $t-t'$ Hubbard models
\cite{Freire,katanin2} yield an (generally speaking,  anisotropic)
suppression of the quasiparticle weight (inverse effective mass) along the Fermi surface,
the staggered spin susceptibility remaining divergent, although the divergence is
considerable weakened.
%suppresses the tendency tohe scale
%remains finite, but considerably suppressed
%indications pointing to the existence of a NFL regime
%at temperature T>0 displaying a truncated Fermi surface
%k-dependence

Note that the picture in the Hubbard model and $s-d(f)$ model (where ``direct''
exchange interaction is present) can be considerably different.
Recently, the $\epsilon$-expansion has been used for a scaling consideration
of the 2D antiferromagnetic Kondo lattice with the use of non-linear sigma model
\cite{2Drg}.

%The situation in the itinerant magnets can be influenced by other factors

In the case of a smooth electron density of states, various physical
properties of NFL systems are discussed within our approach
in Ref. \cite{nfl}. The temperature behavior of
magnetic characteristics $\overline{S}$ and $\overline{\omega }$, which
depend exponentially on the coupling constant, is decisive for the NFL
picture under consideration.
At the same time, the presence of VHS near the Fermi level
influences strongly all the electronic, magnetic and transport properties in
the scaling approach, as well as in the one-electron theory. The replacement
$\rho ^{2}g_{ef}^{2}(T)\rightarrow \rho ^{2}(T)g_{ef}^{2}(T)$ with $\rho
(T)$ being considerably temperature dependent owing to VHS may modify
somewhat the behavior of observable quantities.

Consider the temperature dependence of the magnetic susceptibility $\chi
\propto \overline{S}/\overline{\omega }$. Using the scaling arguments we can
replace $\overline{\omega }\rightarrow \overline{\omega }_{ef}(C),$ $%
\overline{S}\rightarrow \overline{S}_{ef}(C)$ with $|C|\sim T,$ which yields
$\chi (T)\propto T^{-\zeta }$. The non-universal exponent $\zeta $ is
determined by details of magnetic structure (the difference $a-1$ can be
used as a perturbations, see Ref. \cite{nfl}). Besides that, a number of
crossovers are characteristic for NFL behavior under consideration. In the
coherent  \textquotedblleft magnon\textquotedblright\ regime we have $\zeta
=(a-1)/a,$ and in the \textquotedblleft quasilinear\textquotedblright\
(incoherent) region $\zeta =(a-1)A/2$.

The temperature dependence of electronic specific heat can be estimated from
the second-order perturbation theory, $C_{el}(T)/T\propto 1/Z(T)$ where $%
Z(T) $ is the residue of the one-electron Green's function at the distance $T$
from the Fermi level. Then we have
\begin{equation}
C_{el}(T)/T\propto \rho ^{2}(T)g_{ef}^{2}(T)\overline{S}_{ef}(T)/\overline{%
\omega }_{ef}(T)\propto \rho ^{2}(T)g_{ef}^{2}(T)\chi (T)
\end{equation}%
which results in a non-trivial behavior of the Wilson ratio.

Following to Ref.\cite{nfl}, a simple estimation of the transport
relaxation rate (which determines the temperature dependence of the resistivity
owing to scattering by spin fluctuations in AFM phase) yields
\begin{equation}
\ \frac{1}{\tau }\propto T^{2}\rho ^{2}(T)g_{ef}^{2}(T)\overline{S}_{ef}(T)/%
\overline{\omega }_{ex}(T)\propto T^{2}C_{el}(T)/T  \label{taul}
\end{equation}
However, a more refined treatment in spirit of Ref.\cite{Hlubina}
would be useful in some cases.

In Ref.\cite{nfl}, a mechanism of NFL behavior owing to peculiar behavior of
spin spectral function was proposed. Here we treated a similar, but somewhat
more simple and natural mechanism which is connected with the singularities in the bare
electron spectrum.
Of course, a more accurate treatment of magnetic fluctuations near the quantum phase
transition is required.
Therefore detailed investigations of the NFL behavior for a
realistic Fermi surface and spin spectral function are of interest. An
accurate investigation of the situation where VHS is shifted from $E_{F}$
or two peaks are present below and above $E_{F}$ \cite{VKT} would be also
instructive.

The research described was supported in part by the Program
\textquotedblleft Quantum Physics of Condensed Matter\textquotedblright\
from Presidium of Russian Academy of Sciences.
%Grant No.99-02-16279 from?? the Russian Basic Research Foundation.
The author is grateful to M.I. Katsnelson and A.A. Katanin for discussions of
the problem.

\end{document}